\documentclass[twocolumn,aps,prb,showpacs,superscriptaddress]{revtex4}

\usepackage{amsmath}
\usepackage{amssymb}
\usepackage{epsfig}

\renewcommand{\v}[1]{{\bf #1}}

\newcommand{\be}{\begin{equation}}
\newcommand{\ee}{\end{equation}}
\newcommand{\nn}{\nonumber \\}

\newcommand{\ba}{\begin{eqnarray}}
\newcommand{\ea}{\end{eqnarray}}

\newcommand{\gammabar}{\gamma^+}

\newcommand{\bw}{\begin{widetext}}
\newcommand{\ew}{\end{widetext}}

\newcommand{\bpm}{\begin{pmatrix}}
\newcommand{\epm}{\end{pmatrix}}

\newcommand{\bbar}{b^\dag}
\newcommand{\kbar}{\overline{k}}

\begin{document}
\title{Magnetic response of the $J_1 \!-\! J_2$ spin Hamiltonian from \\
classical Monte Carlo and Schwinger boson mean field theory}

\author{Zhihua Yang}
\affiliation{Institute of Basic Science, Sungkyunkwan University,
Suwon 440-746, Korea}
\author{Jung Hoon Kim}
\affiliation{Department of Physics, BK21 Physics Research Division,
Sungkyunkwan University, Suwon 440-746, Korea}
\author{Jung Hoon Han}
\email[Electronic address:$~~$]{hanjh@skku.edu}
\affiliation{Department of Physics, BK21 Physics Research Division,
Sungkyunkwan University, Suwon 440-746, Korea} \email[Electronic
address:$~~$]{hanjh@skku.edu}

\date{\today}

\begin{abstract}
Magnetic susceptibilities at several potential ordering wave vectors
$(0,0)$, $(\pi,0)$, and $(\pi,\pi)$ are analyzed for the
antiferromagnetic $J_1 \!-\! J_2$ spin Hamiltonian by classical
Monte Carlo and Schwinger boson mean field theories over the
parameter range $0\le 2J_2 /J_1 \le 2$. We find a nearly linear-$T$
behavior of the uniform susceptibility that extends up to the
temperature scale of $J_1$ within both calculation schemes when
$2J_2 /J_1$ is sufficiently removed from the critical point $2J_2
/J_1 = 1$. The window of linear temperature dependence diminishes as
the critical point is approached.
\end{abstract}

\maketitle

\section{Introduction}

One of the outstanding issues for the spin density wave (SDW) phase
of the parent FeAs compound is whether a localized spin model can be
justifiably used for the description of spin dynamics. The concern
arises mainly from the fact that, unlike the parent compound of
cuprate high-temperature superconductors, the parent material for
FeAs-based superconductors is metallic and the validity of the
description which ignores the itinerant electrons is not nearly as
water-tight as with the half-filled cuprate compound. Despite these
concerns, the $J_1 \!-\! J_2$ spin model had been proposed for the
spin dynamics of the FeAs parent compound\cite{si,xiang} which,
disregarding the multi-orbital character of the FeAs band structure,
reads

\be H_S = J_1 \sum_{\langle ij \rangle } {\v S}_i \cdot {\v S}_j +
J_2 \sum_{\langle ik \rangle} {\v S}_i \cdot {\v S}_k . \ee
The nearest neighbor (NN) and next-nearest neighbor (NNN) pairs are
written as $\langle ij \rangle$ and $\langle ik \rangle$,
respectively. From first-principles calculations one obtains the
estimate $J_1 \sim J_2$ and $J_1$ in the range of several hundred
Kelvin\cite{yildirim,xiang}.

Of particular relevance in connection with the magnetic response of
FeAs is the behavior of the uniform susceptibility $\chi_u$. The
temperature dependence of $\chi_u$ was found to be linear up to
500-700K\cite{linear-T}, well beyond the SDW ordering temperature
T$_\mathrm{SDW}\sim$ 150K. The relevance of the $J_1 \!-\! J_2$ spin
Hamiltonian to the FeAs phenomenology largely depends on its ability
to reproduce the observed temperature dependence of $\chi_u$. The
uniform susceptibility for the $J_1$-only model has been calculated
semi-classically by Takahashi\cite{takahashi} and more recently by
classical Monte Carlo (MC) calculations\cite{J1-MC}, demonstrating a
roughly linear behavior of $\chi_u (T)$ in the low-temperature
regime below the mean-field transition temperature, T$_\mathrm{MF}$.
The quantum spin-$1/2$ Heisenberg Hamiltonian was also found to have
a low-temperature region with a roughly linear temperature-dependent
$\chi_u$\cite{QMC}.

A recent note by Zhang \textit{et al.}\cite{preformed-moment} cites
such susceptibility behavior as evidence of pre-formed moments in
FeAs in the paramagnetic regime. A quasi-linear temperature
dependence of $\chi_u$ in the $J_1 \!-\! J_2$ spin Hamiltonian was
demonstrated in a spin-wave calculation based on the Dyson-Maleev
transformation of the spin operators\cite{preformed-moment}. A
similar temperature dependence of the susceptibility was also found
in a model based on the nonlinear sigma model description of the
spin dynamics\cite{weng}. Both ideas\cite{preformed-moment,weng}
support, and are based on the presence of pre-formed magnetic
moments in the paramagnetic phase\cite{si,xiang}. In contrast to the
Curie-Weiss susceptibility behavior of a local moment, the
linear-$T$ susceptibility is a reflection of the non-trivial
correlation among the moments in the paramagnetic state, and a
deeper understanding of the magnetic response in such a correlated
paramagnetic phase is highly desirable.

A lot is known about the classical $J_1 \!-\! J_2$ model from
previous theoretical works\cite{CCL,weber}. Although no true
long-range order is possible in two dimensions, the mean-field
ground state is antiferromagnetically ordered at the wave vector $k=
(\pi,\pi)$ when $2J_2 /J_1 <1$. For $2J_2 /J_1>1$, the two
sublattices separately support a mean-field staggered magnetic phase
while the relative orientation of the sublattice magnetization
vectors remains undetermined. It is the thermal fluctuations which
selects a collinear spin order at $k=(\pi,0)$ or $k=(0,\pi)$ through
a order-by-disorder mechanism\cite{CCL}. A quantum spin-$1/2$
version of the $J_1\!-\!J_2$ has recently received some theoretical
attention\cite{capriotti}, but the relevant uniform magnetic
susceptibility $\chi_u$ has not been calculated in the quantum
model.

In this paper, we  re-visit the susceptibility behavior of the $J_1
\!-\! J_2$ spin model in a comprehensive and unified fashion using
both classical Monte Carlo (MC) methods and the Schwinger boson mean
field theory (SBMFT).  Previous analytic and numerical
works\cite{takahashi,J1-MC,QMC,preformed-moment,capriotti} typically
have examined the susceptibility for a particular ordering wave
vector. In the $J_1 \!-\! J_2$ spin model the dominant ordering wave
vector switches from $(\pi,\pi)$ for $2J_2 /J_1 < 1$ to $(\pi,0)$
for $2J_2 /J_1>1$. The fate of the non-dominant susceptibilities,
for instance the susceptibility at $k=(\pi,0)$ for the $2J_2 / J_1
<1$ region, has not been addressed carefully. The uniform
susceptibility $\chi_u$ is not expected to be dominant for any $2J_2
/J_1$ ratio, but its behavior has not been understood in a
systematic way, either.

At the moment the controversy surrounding the
itinerant\cite{korshunov} vs. localized moment
origin\cite{preformed-moment,weng} of the linear-$T$ susceptibility
is an on-going issue. A careful and systematic examination of the
magnetic response of the $J_1 \!-\! J_2$ spin model by employing
well-established techniques is expected to be a helpful addition to
our knowledge required to settle the issue in the long run. The
organization of the paper is as follows. In Sec.~\ref{sec:MC}, Monte
Carlo calculation of the magnetic susceptibilities at several wave
vectors are presented for all $2J_2/J_1$ ratios between 0 and 2,
which encompasses the Heisenberg limit $2J_2 /J_1 = 0$ and the
critical point at $2J_2 /J_1 = 1$. And in Sec.~\ref{sec:SBMFT} a
Schwinger boson mean-field calculation employing several order
parameters which embody both $(\pi,\pi)$ and $(\pi,0)$ ordering are
presented over the same $2J_2 /J_1$ range. We conclude with a
discussion in Sec.~\ref{sec:summary}. The properties of the uniform
and $(\pi,\pi)$-staggered magnetic susceptibilities of the
$J_1$-only Hamiltonian in the SBMFT are reviewed in the Appendix.

\section{Monte Carlo Evaluation of the Magnetic Susceptibility}
\label{sec:MC}

\begin{figure}
\begin{center}
\includegraphics[scale=1.1]{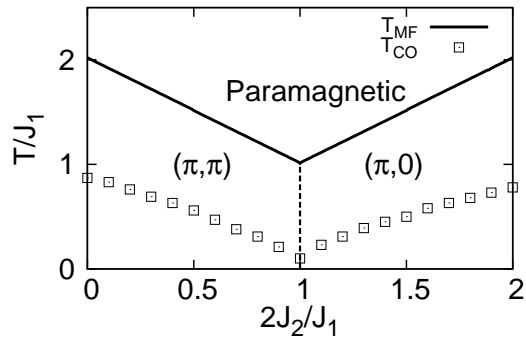}
\end{center}
\caption{Mean-field phase diagram of the $J_1 \!\!-\!\! J_2$
classical spin model. $(\pi,\pi)$-ordered phase is separated by a
first-order line (dotted) at $2J_2 /J_1 =1$ from the
$(\pi,0)$-ordered phase. Crossover temperature T$_\mathrm{CO}$
extracted from the uniform susceptibility in the classical Monte
Carlo calculation is shown for comparison. }
\label{fig:phase-diagram}
\end{figure}

Classical mean field analysis of the $J_1 \!-\! J_2$ model assuming
the ordering pattern $\langle S_i \rangle = m_1 (-1)^{x_i} + m_2
(-1)^{x_i + y_i}$ (coexistence of $(\pi,0)$ and $(\pi,\pi)$ order)
is reduced to the coupled set of mean-field equations

\ba m_1 \!+\! m_2 \!&=&\! \tanh \!\left( \frac{2}{T} [(J_2 m_1 \!+\!
(J_1 \!\!-\!\! J_2) m_2 ]\right) , \nn
m_1 \!-\! m_2 \!&=&\! \tanh \!\left( \frac{2}{T} [J_2 m_1 \!-\! (J_1
\!\!-\!\! J_2) m_2 ]\right). \ea
As shown in Fig. \ref{fig:phase-diagram}, either a
$(\pi,\pi)$-ordered phase ($m_1 = 0, m_2 \neq 0$) or a
$(\pi,0)$-ordered phase ($m_1 \neq 0, m_2 = 0$) is favored depending
on the $2J_2 /J_1$ value being less or greater than one. No
co-existence region exists where $m_1$ and $m_2$ are simultaneously
nonzero. A first-order line separates the two phases at $2J_2 /J_1
=1$. In the Ginzburg-Landau picture, a repulsive interaction $v |m_1
|^2 |m_2 |^2$ term with $v>0$ exists between the order parameters.

With this classical mean field phase diagram in mind, we have
measured the magnetic susceptibilities at three wave vector
positions, $k=(0,0), (\pi,0)$, and $(\pi,\pi)$, for the range $0 \le
2J_2 /J_1 \le 2$ that also encompasses the critical point $2J_2 /J_1
= 1$. For a given $k$ vector, the susceptibility $\chi^\alpha_k$ for
the component $\alpha=x,y,z$ is computed according to

\be \langle S^\alpha_k \rangle \!=\! \sum_i \langle S^\alpha_i
\rangle e^{-ik\cdot r_i}, ~
\chi^\alpha_k \!=\! { \frac{\langle S^\alpha_{-k} S^\alpha_k \rangle
\!-\! \langle S^\alpha_{-k} \rangle \langle S^\alpha_k \rangle}{T N}
}. \label{eq:chi-k}\ee
Here $N$ stands for the number of lattice sites, and $\langle \cdots
\rangle$ is the thermal average.

In agreement with the mean-field analysis, the $(\pi,\pi)$
susceptibility rises sharply at a finite temperature $T_c$ in the
region $2J_2 /J_1 <1$. For the $(\pi,0)$ susceptibility the sharp
rise occurs on the $2J_2 /J_1 > 1$ side. At $2J_2 /J_1 = 1$ both
susceptibilities showed increasing behavior at lower temperatures.
This temperature $T_c$ however decreases when a larger lattice size
is used in the simulation, in accordance with the Mermin-Wagner
theorem which prohibits long-range order in a continuous spin model.
We will not discuss the behavior of the dominant susceptibilities
any further and turn our attention to non-dominant susceptibilities.

\begin{figure}
\begin{center}

\begin{tabular}{cc}
     \resizebox{40mm}{!}{\includegraphics{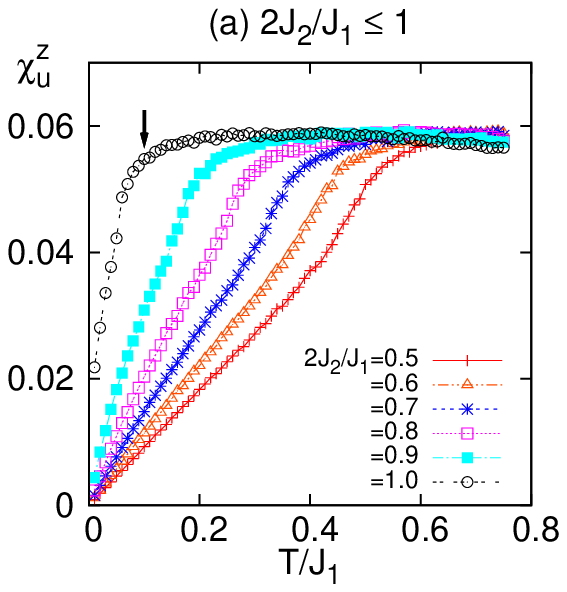}} &
     \resizebox{40mm}{!}{\includegraphics{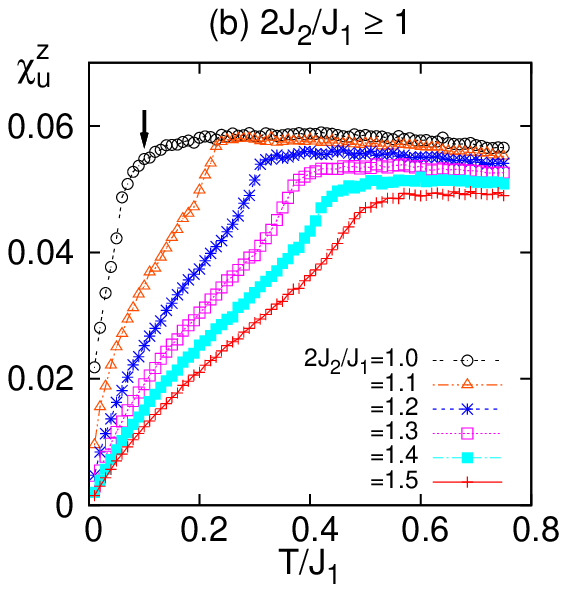}} \\
     \resizebox{40mm}{!}{\includegraphics{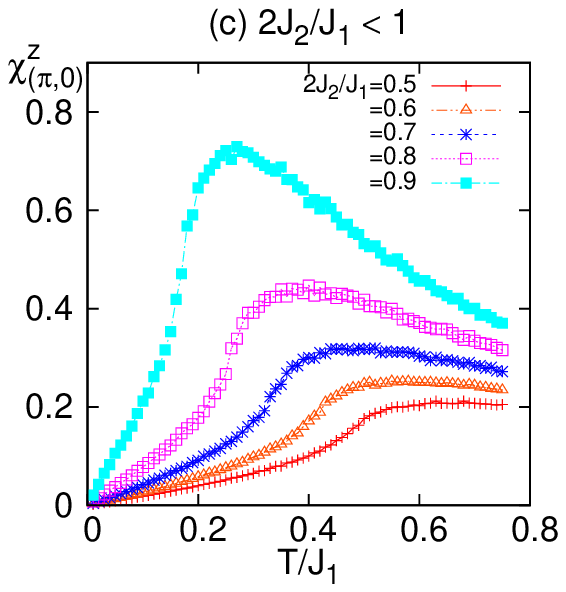}} &
     \resizebox{40mm}{!}{\includegraphics{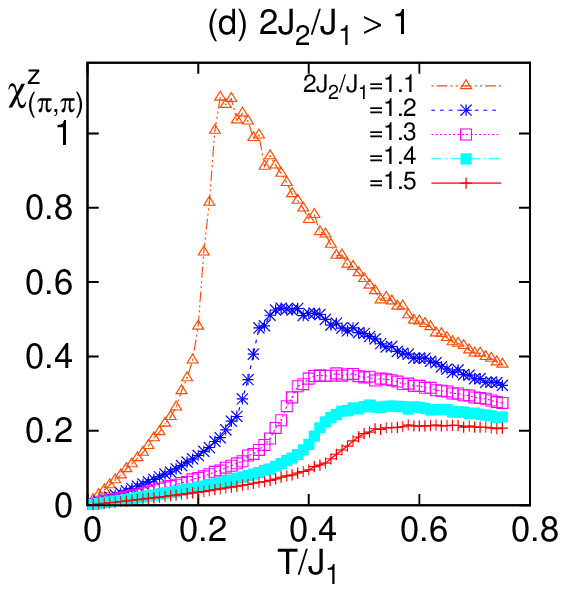}} \\
     \resizebox{37mm}{!}{\includegraphics{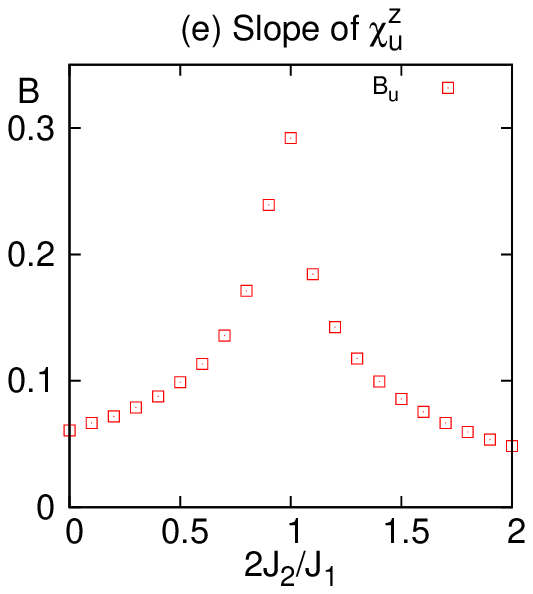}} &
     \resizebox{37mm}{!}{\includegraphics{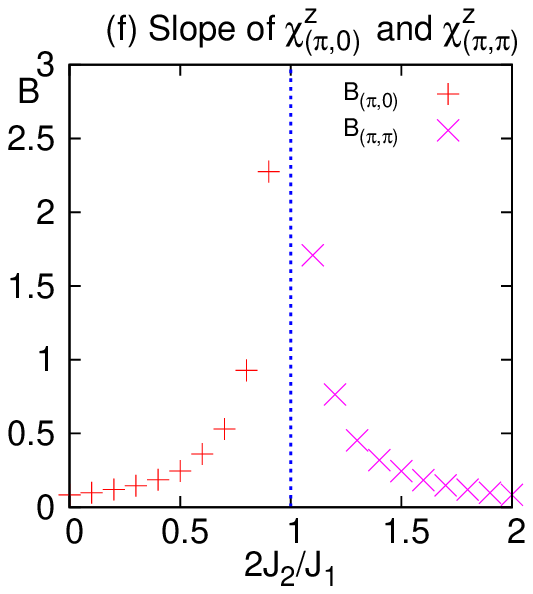}} \\
  \end{tabular}

\end{center}
\caption{(color online) Temperature dependence of (a) uniform
susceptibility $\chi^z_{(0,0)}$ for $2J_2 / J_1 \le 1$, (b)
$\chi^z_{(0,0)}$ for $2J_2 /J_1 \ge 1$, (c) staggered susceptibility
$\chi^z_{(\pi,0)}$ for $2J_2 /J_1 < 1$, and (d) staggered
susceptibility $\chi^z_{(\pi,\pi)}$ for $2J_2 /J_1 > 1$. Data are
obtained for a 16$\times$16 lattice with $5\times 10^5$ MC steps per
data point. Several calculations run on a 40$\times$40 lattice gave
virtually identical curves as those of a 16$\times$16 lattice over
the whole temperature interval. The crossover temperature such as
indicated by an arrow in (a) and (b) marks the deviation from
low-temperature linearity. We have plotted the $z$-component of the
susceptibilities for an easier determination of
$T_\mathrm{CO}$\cite{details-of-MC}. (e) Slopes of the
low-temperature linear part of $\chi^z_{(0,0)}$ vs. $T/J_1$. (f)
Slopes of the low-temperature linear part of $\chi^z_{(\pi,0)}$ for
$2J_2 /J_1 < 1$ and $\chi^z_{(\pi,\pi)}$ for $2J_2 /J_1 > 1$. Both
slopes become sharper near the critical point.} \label{fig:suscep}
\end{figure}

Results of the MC calculation for the non-divergent susceptibilities
are shown in Fig. \ref{fig:suscep}. The uniform susceptibility
$\chi_{(0,0)}$, which remains non-divergent for all values of $2J_2
/J_1$, does show the low-temperature linear-$T$ behavior up to a
certain crossover temperature scale,
$T_\mathrm{CO}$\cite{details-of-MC}. Furthermore, $T_\mathrm{CO}$
grows as the $2J_2 \!-\! J_1$ value moves away from criticality. The
crossover temperature $T_\mathrm{CO}$ can be extracted from the
$\chi_{(0,0)}$ data due to the fairly abrupt deviation from
low-temperature linearity at $T\approx T_\mathrm{CO}$. A plot of
$T_\mathrm{CO}$ over the range $0 \le 2J_2 /J_1 \le 2$ is shown in
Fig. \ref{fig:phase-diagram} along with the mean-field transition
temperature $T_\mathrm{MF}$. Due to the thermal fluctuation effects,
$T_\mathrm{CO}$ remains substantially smaller than the corresponding
$T_\mathrm{MF}$. Even at the critical point $2J_2/J_1 =1$ the
crossover scale remains finite at $T_\mathrm{CO}/J_1 \sim 0.1$.

The behavior of non-divergent $\chi_{(\pi,0)}$ for $2J_2 /J_1 < 1$,
and $\chi_{(\pi,\pi)}$ for $2J_2 /J_1 > 1$ are also shown in Fig.
\ref{fig:suscep}. Interestingly, these susceptibilities also show a
linear temperature dependence at low temperatures, and find the
maximum value at $T \approx T_\mathrm{CO}$. Unlike the uniform
susceptibility, these two quantities exhibit divergent behavior right
at the critical point $2J_2 /J_1 \!=\! 1$, and are omitted from the
plot in Fig. \ref{fig:suscep}. In terms of absolute scale, the
$(\pi,0)$ and $(\pi,\pi)$ susceptibilities are about an order of
magnitude bigger than $\chi_{u}$. The linearity can be characterized
by fitting the low-temperature part with the form $\chi^z = B
(T/J_1)$. The slope $B$ of the linear parts of the susceptibilities
are shown in Fig. \ref{fig:suscep}(e) for the uniform part and Fig.
\ref{fig:suscep} for the staggered parts.

Based on the results of MC calculations, we conclude that the
linear-$T$ behavior of the uniform susceptibility is a natural
aspect of the $J_1 \!-\! J_2$ spin model for the $2J_2 /J_1$ ratio
that spans both sides of the critical point, and presumably for a
wider class of spin models whose mean-field ordering wave vector
occurs well away from $(0,0)$. In fact, the linearity of $\chi_u
(T)$ persists over a wider temperature range when $2J_2 /J_1$
departs further from the critical value. If we take the estimate
$J_1 = J_2 = 500$K\cite{xiang}, the expected crossover temperature
from the MC calculation is $T_\mathrm{CO}\sim 0.8 J_1 \sim 400$K, in
fair proximity to the 500-700 K up to which the linear
susceptibility has been observed. It is also clear that the
linear-$T$ susceptibility behavior alone does not imply the
proximity of the FeAs system to a critical point nor the existence
of high degree of spin frustration.

\section{Schwinger Boson Mean Field Theory of the Magnetic Susceptibility}
\label{sec:SBMFT}

The quantum analogue of the classical susceptibilities can be
obtained by treating the $J_1\! -\! J_2$ spin Hamiltonian in the
framework of Schwinger boson mean field theory (SBMFT). A prior
calculation\cite{preformed-moment} assumed an ordering wave vector
$K_x = (\pi,0)$, or equivalently at $K_y = (0,\pi)$, and ignored the
possibility of ordering at $Q = (\pi,\pi)$. While this assumption
may be sound deep in the $J_2$-dominated part of the phase diagram,
here we want to make an unbiased statement that spans both small and
large $2J_2 /J_1$ regions and keep both $(\pi,0)$ and $(\pi,\pi)$
order parameters in the theory.  The details of the self-consistent
Schwinger boson calculation for the $J_1$-only spin model can be
found in the work of Auerbach and Arovas\cite{AA}.

\begin{figure}[ht]
\begin{center}
\includegraphics[width=0.8\columnwidth]{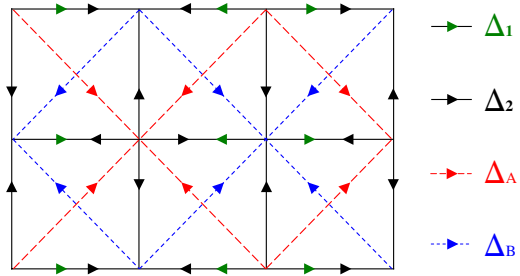}
\end{center}
\caption{(color online) Schematic picture of the pairing amplitudes
$\Delta_{ij}$ used in our calculation. The values of $\Delta_{ij}$
for the arrows going from site $i$ to site $j$ are indicated on the
right. Note that $\Delta_{ji} = -\Delta_{ij}$. } \label{fig:delta}
\end{figure}

For SBMFT calculations one first re-writes the $J_1\! -\! J_2$
Hamiltonian in the mean-field form

\ba && H_{MF} = - \frac{J_1}{2} \sum_{\langle ij\rangle}
\Delta_{ij}^* A_{ij} - \frac{J_2}{2} \sum_{\langle ik \rangle}
\Delta_{ik}^* A_{ik} + h.c. \nn
&& ~~~~~ +\! \sum_i \lambda_i \Bigl(\bbar_{i1} b_{i1} + \bbar_{i2}
b_{i2}\!-\! 2S \Bigr), \label{eq:H-SBMFT}\ea
where the operators are $A_{ij} = b_{i1}b_{j2} - b_{i2}b_{j1} =
-A_{ji}$, and $\Delta_{ij} =\langle A_{ij}\rangle $. The $b_{i1}$ and
$b_{i2}$ are the two species of boson operators that obey the
occupation constraint $\bbar_{i1} b_{i1} + \bbar_{i2} b_{i2}\!=\!
2S$, which is enforced with the Lagrange multiplier $\lambda_i$.

The distribution of $\Delta_{ij}$ adopted in our calculation is
schematically shown in Fig.~\ref{fig:delta}. The nearest-neighbor
pairing fields are assumed to be modulated at both $K_x$ and $Q$
with the respective amplitudes $\Delta_{1}$ and $\Delta_2$:

\ba &&\Delta_{i,i\pm\hat{x}}=\Delta_2 (-1)^{x_i + y_i} + \Delta_{1x}
(-1)^{x_i}, \nn &&\Delta_{i,i\pm\hat{y}}=\Delta_2 (-1)^{x_i + y_i}+
\Delta_{1y} (-1)^{x_i} . \ea
By symmetry, it turns out, $\Delta_{1y}$ vanishes identically and we
can re-write $\Delta_{1x} = \Delta_1$. For the assumed ordering at
$(0,\pi)$ we would have $\Delta_{1x}=0$ instead. Without loss of
generality we choose to work with the $(\pi,0)$ modulation below. As
for the interactions among the next-nearest neighbors let us write

\ba \Delta_{i,i\pm\hat{x}\pm\hat{y}}=\frac{\Delta_A \! + \! \Delta_B
}{2}(-1)^{y_i}+ \frac{\Delta_A \!-\! \Delta_B }{2}(-1)^{x_i} . \ea
It will be shown that $\Delta_A = \Delta_B$ whenever they are
nonzero.

Due to the enlarged unit cell in real space assumed by the ordering
pattern, the Brillouin zone is reduced to
$[-\frac{\pi}{2},\frac{\pi}{2}]\times[-\frac{\pi}{2},\frac{\pi}{2}]$,
and the Hamiltonian is expressed with an 8-component spinor
%

\ba && \psi_k \!=\! \bpm b_{k1}  \\ b_{k\!+\!K_x 1} \\ b_{k\!+\!K_y
1} \\ b_{k\!+\!Q 1} \\ \bbar_{\kbar 2}\\   \bbar_{\kbar \!+\! K_x
2}\\  \bbar_{\kbar \!+\! K_y 2}\\   \bbar_{\kbar\!+\! Q 2} \epm , ~~
H = \sum'_k \psi^+_k\bpm \lambda & {\cal H}_k^+ \\ {\cal H}_k &
\lambda \epm\psi_k .\ea

The 4$\times$4 matrix ${\cal H}_k$ is given by

\ba \bpm
0   & -\Delta_{1k}\!-\!\Delta_{3k}& -\Delta'_{3k} & -\Delta_{2k}\\
\Delta_{1k}\!+\!\Delta_{3k} & 0 &\Delta'_{2k} & \Delta'_{3k}\\
\Delta'_{3k} & -\Delta'_{2k}& 0 & -\Delta_{1k}\!+\!\Delta_{3k}\\
\Delta_{2k}& -\Delta'_{3k} & \Delta_{1k} \!-\!\Delta_{3k} & 0\epm,
\ea
where

\ba \label{eq:elements}
\Delta_{1k}&=&J_1\Delta_{1}\cos k_x,\nn
%
%
\Delta_{2k}&=&J_1\Delta_2(\cos k_x\!+\!\cos k_y), \nn
\Delta'_{2k}&=&J_1\Delta_2(\cos k_x\!-\!\cos k_y),\nn
\Delta_{3k}&=& J_2(\Delta_A\!-\!\Delta_B)\cos k_x\cos k_y ,\nn
\Delta'_{3k}&=& J_2(\Delta_A\!+\!\Delta_B)\cos k_x\cos k_y .
\label{eq:Delta-k}\ea
Note the antisymmetry $({\cal H}_k )_{\mu\nu} = -({\cal H}_k
)_{\nu\mu}$. The sum $\sum'_k$ extends over the reduced Brillouin
zone (RBZ) only. The four kinds of momentum ``flavors" ($k,
k\!+\!K_x, k\!+\!K_y, k\!+\!Q$) can be organized with an index
$\alpha =1\sim4$: $(b_{k \sigma}, b_{k\!+\!K_x \sigma}, b_{k\!+\!K_y
\sigma}, b_{k\!+\!Q \sigma}) \! \rightarrow \! (b_{k, 1 \sigma},
b_{k, 2 \sigma}, b_{k, 3 \sigma}, b_{k, 4 \sigma})$. Rotation to the
eigenoperators are implemented through

\ba b_{k, \alpha 1} &=& \sum_{a=1}^4 \Bigl( u_{k, \alpha a}
\gamma_{k, a1} + v^*_{k, \alpha a} \gammabar_{k, a2} \Bigr), \nn
 b_{\kbar, \alpha 2} &=& \sum_{a=1}^4 \Bigl( u_{k, \alpha a} \gamma_{k, a2} -
v^*_{k, \alpha a} \gammabar_{k, a1} \Bigr) , \label{eq:expansion}\ea
which would yield

\be H = \sum'_{k,a} E_{k,a}\left( \gamma^+_{k, a1} \gamma_{k, a1}
\!+\! \gamma^+_{k, a2} \gamma_{k, a2} \right). \ee
It is convenient to group the wave function components as $\v u_{k,
a} = (u_{k, 1a}, u_{k, 2a},u_{k, 3a}, u_{k, 4a})^T$, and $\v v_{k,
a} = (v_{k, 1a}, v_{k, 2a}, v_{k, 3a}, v_{k, 4a})^T $, after which
the equations of motion becomes

\ba  \lambda \v u_{k, a} + {\cal H}^+_k \v v_{k, a} &=& E_{k, a} \v
u_{k, a}, \nn \lambda \v v_{k, a} + {\cal H}_k \v u_{k, a} &=&
-E_{k, a} \v v_{k,a} , \label{eq:non-Hermitian}\ea
%
%
or
\ba  (\lambda^2 \!-\! E_{k, a}^2)  \v u_{k, a} &=& {\cal H}^+_k
{\cal H}_k \v u_{k, a}, \nn (\lambda^2 \!-\! E_{k, a}^2)  \v v_{k,
a} &=& {\cal H}_k {\cal H}^+_k \v v_{k, a} . \label{eq:u-and-v}\ea
The matrix ${\cal M}_k ={\cal H}^+_k {\cal H}_k={\cal H}_k {\cal
H}^+_k$ is Hermitian and all its eigenvalues are non-negative. It
proved to be an enormous advantage to deal with the ordinary
Hermitian matrix problem of Eq. (\ref{eq:u-and-v}) over a
non-Hermitian one posed by Eq. (\ref{eq:non-Hermitian}). With the
eigenvalues of ${\cal M}_k$ written $m_{k,a}^2$, the energy spectrum
is obtained as $E_{k,a} = \sqrt{\lambda^2 - m_{k,a}^2}$.

One can either solve for $\v u_{k, a}$ or $\v v_{k, a}$ from Eq.
(\ref{eq:u-and-v}), and derive the other part using

\be \v v_{k,a} =  -\frac{{\cal H}_k \v u_{k, a} }{ E_{k,a} \!+\!
\lambda} ~~ \mathrm{or} ~~ \v u_{k,a} =  \frac{{\cal H}^+_k \v v_{k,
a}}{ E_{k,a} \!-\! \lambda} . \ee
The normalization for $\v u_{k,a}$ and $\v v_{k,a}$ reads

\be \v u^*_{k, a} \cdot \v u_{k, b} = \frac{\lambda \!+\!E_{k,a}}{
2E_{k,a}}\delta_{ab}, ~~ \v v^*_{k, a} \cdot \v v_{k, b} =
\frac{\lambda \!-\! E_{k,a}}{ 2E_{k,a}}\delta_{ab}, \ee
to ensure $\v u^*_{k, a} \cdot \v u_{k, b} \!-\! \v v^*_{k, a} \cdot
\v v_{k, b} \!=\! \delta_{ab}$.

The averages of the boson operators we need are

\be \langle b_{k, \alpha 1} b_{\kbar, \beta 2} \rangle \!=\!
\sum_{a=1}^4 \Bigl(v^*_{k, \alpha a} u_{k, \beta a} B_{k, a}
\!-\!u_{k, \alpha a} v^*_{k, \beta a} (1\!+\! B_{k,a})\Bigr) \ee
where $B_{k,a}$ is the Bose-Einstein function of energy $E_{k, a}$,
and $\kbar = -k$. The order parameters introduced for the mean field
calculation are

\begin{widetext}

\ba \Delta_{1}&=&-\frac{2}{N} \sum_k \cos k_x \langle b_{k1}
b_{\kbar \!+\!K_x 2} \rangle \nn
&=& \frac{2}{N}\sum_{k,a}' \cos k_x (u_{k,1a}v_{k,2a}^*  \!-\!
u_{k,2a}v_{k,1a}^* \!+\!u_{k,3a}v_{k,4a}^*\!-\! u_{k,4a}v_{k,3a}^* )
\coth\left( \frac{\beta E_{k,a}}{ 2}\right), \nn
\Delta_2&=&-\frac{1}{N} \sum_k (\cos k_x \!+\!\cos k_y)\langle
 b_{k1} b_{\kbar\!+\!Q 2} \rangle  \nn
&=& \frac{1}{N}\sum_{k,a}' \left[(\cos k_x \!+\!\cos
k_y)(u_{k,1a}v_{k,4a}^* \!-\! u_{k,4a}v_{k,1a}^*)\!+\!(\cos k_x
\!-\! \cos k_y)(u_{k,3a}v_{k,2a}^* \!-\!
u_{k,2a}v_{k,3a}^*)\right]\coth\left( \frac{\beta E_{k,a}}{
2}\right) , \nn
\Delta_A&=&-\frac{2}{N} \sum_k \cos k_x \cos k_y \langle b_{k1}
b_{\kbar + K_y 2}+ b_{k1} b_{\kbar + K_x 2}  \rangle \nn
&=&\frac{2}{N}\sum_{k,a}' \cos k_x\cos k_y [(u_{k,1a}\!+\!
u_{k,4a})(v_{k,2a}^*+v_{k,3a}^*)\!-\!(u_{k,2a}\!+\! u_{k,3a})
(v_{k,1a}^* \!+\!v_{k,4a}^*)]\coth\left( \frac{\beta E_{k,a}
}{2}\right), \nn
\Delta_B&=&\frac{2}{N} \sum_k \cos k_x \cos k_y \langle b_{k1}
b_{\kbar + K_x 2} - b_{k1} b_{\kbar  + K_y 2} \rangle  \nn
&=& \frac{2}{N}\sum_{k,a}' \cos k_x\cos
k_y[(u_{k,2a}\!-\!u_{k,3a})(v_{k,1a}^*\!-\!v_{k,4a}^*)\!-\!
(u_{k,1a}\!-\!u_{k,4a})(v_{k,2a}^* \!-\! v_{k,3a}^* )]\coth\left(
\frac{\beta E_{k,a}}{2}\right),   \nn
2S\!+\!1 &=& \frac{1}{N}\sum_{k,a}' \frac{\lambda}{E_{k,a}}
\coth\left( {\beta E_{k,a} \over 2}\right) . \ea
\end{widetext}
The last line reflects the boson occupation numbers satisfy $(1/N)
\sum_k \langle \bbar_{k,1} b_{k,1} +\bbar_{k,2} b_{k,2} \rangle =
2S$.

\begin{figure}
\begin{center}
\includegraphics[scale=1.1]{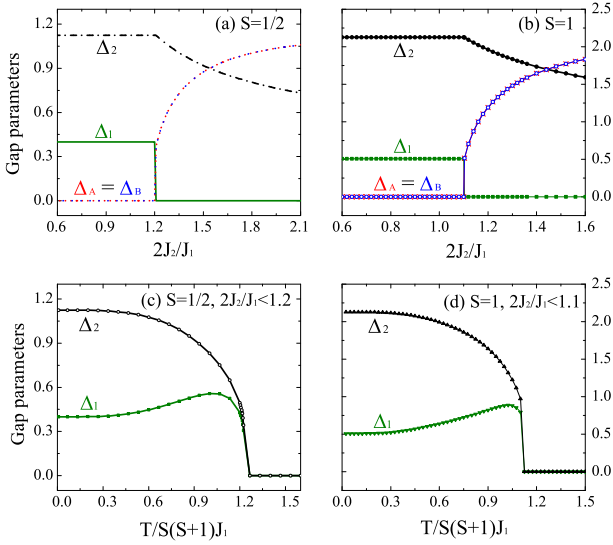}
\end{center}
\caption{(color online) Dependence of the gap parameters on the
ratio $2J_2 /J_1$ at low temperature, $T/J_1 =0.001$ for (a) $S=1/2$
and (b) $S=1$. Temperature dependence of the gap parameters
$\Delta_1$ and $\Delta_2$ when $2J_2 /J_1 < \eta_c$ for (c) $S=1/2$
and (d) $S=1$. The temperature is divided by the exchange energy
scale $J_1 S(S+1)$. The plot shows the onset of both $\Delta_1$ and
$\Delta_2$ at the same temperature $T^\mathrm{SBMFT}_\mathrm{MF}$. }
\label{fig:gap-parameter}
\end{figure}

We have considered two spin values, $S=1/2$, and $S=1$, in the
self-consistent calculation. The low-temperature ($T/J_1 =0.001$)
behavior of the gap parameters are shown in Fig.
\ref{fig:gap-parameter} over the range $0\le 2J_2 /J_1 \le 2$. There
are two regions showing distinct behaviors, separated at $2J_2
/J_1=\eta_c$, $\eta_c$ is  a spin-dependent critical point. For
$S=1/2$ the critical point is $\eta_c=1.2$, but $\eta_c=1.1$ when
$S=1$\cite{capriotti}. We found that $\eta_c=1$ as long as $S\ge 2$,
which is in agreement with the classical result. In Figs.
\ref{fig:gap-parameter} (a) and (b), for $2J_2 /J_1 < \eta_c$ one
finds $\Delta_A = 0 =\Delta_B$, and $\Delta_1 \neq 0 \neq \Delta_2$,
and $2J_2 /J_1
> \eta_c$ gives $\Delta_A = \Delta_B$ nonzero, but $\Delta_1 = 0$. On
the other hand $\Delta_2$ remains nonzero throughout the whole
range, although its magnitude decreases with the onset of nonzero
$\Delta_A$.

A surprising feature of the SBMFT result that needs to be emphasized
is that $\Delta_1$, representing $(\pi,0)$ order for the NN bonds in
the $x$-direction, becomes nonzero even at $2J_2 /J_1 = 0$ and stays
finite (constant, in fact) all through $2J_2 /J_1 =\eta_c$. Their
temperature dependence for $2J_2 /J_1 < \eta_c$ is shown in Fig.
\ref{fig:gap-parameter} (c) and (d), which tell us that both gap
parameters become non-zero at the same temperature. A spontaneous
loss of rotational symmetry is implied by nonzero $\Delta_1$. As
long as $\Delta_A = 0 = \Delta_B$ the $J_1 \!-\! J_2$ Hamiltonian in
SBMFT is the same as the $J_1$-only model, with no influence from
the presence of the $J_2$ interaction. This explains the constancy
of the gap parameters throughout $0 \le 2J_2 /J_1 < \eta_c$.

\begin{figure}
\includegraphics[width=1.0\columnwidth]{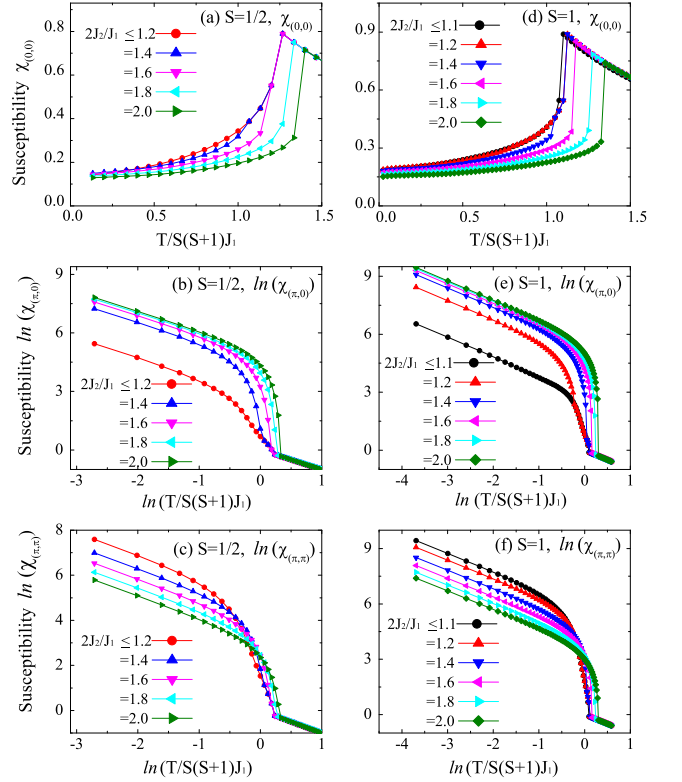}
\caption{(color online) Temperature dependence of the
susceptibilities for several $2J_2 /J_1$ ratios for $S=1/2$ (left
column) and $S=1$ (right column). (a) and (d): $\chi_{(0,0)}$, (b)
and (e) $\chi_{(\pi,0)}$. (c) and (f): $\chi_{(\pi,\pi)}$.
$\chi_{(\pi,0)}$ and $\chi_{(\pi,\pi)}$ are shown on the log-log
scale to emphasize the $1/T$ divergence at low temperature.}
\label{fig:sus-SBMFT}
\end{figure}

The equal-time magnetic susceptibility at wave vector $k$ given in
Eq. (\ref{eq:chi-k}) can be calculated, where $\langle \v S_k
\rangle = 0$ and thus

\ba \chi_k \!=\! { \langle \v S_{\kbar} \cdot\! \v S_k \rangle \over
T N } \!=\! {3\over 4 TN} \sum_{i,j} \Bigl( \langle S^+_i S^-_j
\rangle \!+\! \langle S^-_i S^+_j \rangle \Bigr) e^{ik \cdot (r_i
\!-\! r_j )} . \nn\label{eq:chi-k-again}\ea
The SU(2) invariance of the spin-spin correlation was assumed in
eliminating the $\langle S_i^z S_j^z \rangle$ correlation from the
above. The structure factor $S^+_k = \sum_{i,j}\langle S^+_i S^-_j
\rangle e^{ik\cdot (r_i - r_j )}$ in SBMFT reads

\ba && S^+_k \!=\! \sum_{k_1, k_2} \Bigl( \langle \bbar_{k_1 ,
1}b_{k_1 \!-\! k , 2} \rangle \langle \bbar_{k_2, 2} b_{k_2 \!+\!
k,1} \rangle  + \nn
&&  \langle \bbar_{k_1, 1} \bbar_{k_2, 2}\rangle \langle b_{k_1
\!-\! k, 2} b_{k_2 \!+\! k, 1} \rangle \!+\! \langle \bbar_{k_1, 1}
b_{k_2, 1}\rangle \langle b_{k_1 \!-\! k, 2} \bbar_{k_2 \!-\! k, 2}
\rangle \Bigr) . \nn
\label{eq:S+k}\ea
The first term on the right-hand side, corresponding to the spin
flip process, is identically zero. The other structure factor
$S^-_{k} = \sum_{i,j} \langle S^-_i S^+_j \rangle e^{ik \cdot (r_i -
r_j ) }$ is obtained from $S^-_k = S^+_{-k}$, and the magnetic
susceptibility reads

\be \chi_k = {1\over 2 TN} \Bigl(S^+_k + S^+_{-k} \Bigr).
\label{eq:chi-k-SBMFT}\ee
A factor $3/2$ has been divided out in arriving at Eq.
(\ref{eq:chi-k-SBMFT}) in order to obtain the correct
high-temperature results\cite{AA}.

Three susceptibilities, $\chi_{(0,0)}$, $\chi_{(\pi,0)}$, and
$\chi_{(\pi,\pi)}$, for spin $S=1/2$ and $S=1$, were calculated as
shown in Fig. \ref{fig:sus-SBMFT} for several $2J_2 /J_1$ ratios.
Above the mean-field transition temperature
T$^\mathrm{SBMFT}_\mathrm{MF}$, all the gap parameters vanish
identically and all the susceptibilities obey the Curie-Weiss form
$\chi_k (T) = J_1 S(S+1) /T$ irrespective of $k$. Due to the
constancy of the gap parameters for $2J_2 /J_1 \le \eta_c$ there are
no variations in the susceptibility in this parameter range and it is
sufficient to show only one susceptibility from this region in Fig.
\ref{fig:sus-SBMFT}. We find T$^\mathrm{SBMFT}_\mathrm{MF}$ depends
on the spin $S$ in the manner T$^\mathrm{SBMFT}_\mathrm{MF}\sim t J_1
S(S+1)$ where $t$ is a $2J_2/J_1$-dependent number. The staggered
susceptibilities are plotted on a log-log scale in Fig.
\ref{fig:sus-SBMFT} to emphasize the $1/T$ divergence at low
temperatures.

In the Appendix we show that the uniform susceptibility in the $J_1$
model follows the form $\chi_{(0,0)} = A + B T$ at low temperatures.
Clearly the features of the uniform susceptibilities for all ranges
of $2J_2 /J_1$ are similar, with the main difference lying in the
values of $T^\mathrm{SBMFT}_\mathrm{MF}$ at which the collapse of the
gap parameters occur and the susceptibility follows the Curie-Weiss
form. The reduction in the slope $B$ (defined in a loose sense as the
amount of increase divided by the temperature interval) with
increasing $2J_2 /J_1$ is found in both SBMFT and MC calculations.
Although not nearly linear, the monotonic increase of the uniform
susceptibility occurs over a larger temperature window when $2J_2
/J_1$ is further removed from the critical point, in agreement with
the earlier findings of Monte Carlo calculation.

The two staggered susceptibilities $\chi_{(\pi,0)}$ and
$\chi_{(\pi,\pi)}$ show the temperature dependence which are
qualitatively similar for all $2J_2/J_1$ ranges studied and thus
bears strong resemblance to that of $\chi_{(\pi,\pi)}$ in the $J_1$
model. The properties of the latter quantity are discussed in the
Appendix. Both susceptibilities diverge at low temperature as

\be \chi_{(\pi,0)} \sim {N \over T}, ~~\mathrm{and} ~~
\chi_{(\pi,\pi)} \sim {N \over T} \ee
on both sides of the critical ratio $\eta_c$. The reason for the
appearance of the factor $N$ is explained in the Appendix as well.
The reason for the simultaneous divergence of the two
susceptibilities is the presence of $\Delta_1$ for $2J_2 /J_1 <
\eta_c$ and $\Delta_{A(B)}$ for $2J_2 /J_1 > \eta_c$, either of which
would imply (quasi) ordering at $k=(\pi,0)$. Due to the impossibility
of long range ordering in two dimensions, the associated magnetic
susceptibilities continue to diverge down to zero temperature. The
scaling of the susceptibility with $N$ is tied to the near
Bose-Einstein condensation at low temperature (because true
Bose-Einstein condensation cannot occur in two dimensions, either) as
also discussed in the Appendix for the $J_1$ model. Finally,
$\chi_{(0,\pi)}$ grows as $\sim 1/T$ without the factor $N$, and
numerically much smaller than either $\chi_{(\pi,0)}$ or
$\chi_{(\pi,\pi)}$.

The $(0,\pi)$ order, which has the corresponding gap parameter equal
to zero, plays the role of the non-dominant susceptibilities in
SBMFT. The Monte Carlo calculation in the previous section showed
that such non-dominant susceptibilities, \textit{i.e.}
$\chi_{(\pi,0)}$ and $\chi_{(0,0)}$ for $2J_2 /J_1 < 1$ and
$\chi_{(\pi,\pi)}$ and $\chi_{(0,0)}$ for $2J_2 /J_1 > 1$, are also
non-divergent, going down to a constant value at zero temperature.
Although $\chi_{(0,0)}$ behaves in the similar way in both
treatments, $\chi_{(0,\pi)}$ behaves quite differently in SBMFT
(diverging like $1/T$) as it does in classical theories (going down
to a constant value). Even when a larger spin $S$ is used in SBMFT,
all three staggered susceptibilities diverge at low temperature and
it is not obvious if the classical MC results can be fully recovered
within the SBMFT. Regrettably, little is known in the literature
about the connection of the SBMFT in the large-$S$ limit to the
classical results of the same Hamiltonian. While this would be an
interesting issue in its own right, we leave it as a subject of
future consideration.

The findings of the SBMFT study can be summarized as follows. The
uniform susceptibility shows a monotonic temperature dependence that
can be roughly understood as linearity. The mean-field transition
temperature $T^\mathrm{SBMFT}_\mathrm{MF}$ plays a role similar to
the crossover temperature $T_\mathrm{CO}$ in the classical theory,
and has a larger value when $2J_2/J_1$ gets larger (but only above
the critical value $\eta_c$). Secondly, the two staggered
susceptibilities $\chi_{(\pi,\pi)}$ and $\chi_{(\pi,0)}$ show the
divergent behavior which is consistent with that of
$\chi_{(\pi,\pi)}$ in the $J_1$ model. Both $(\pi,\pi)$ and $(\pi,0)$
(quasi) ordering are present in the SBMFT calculation for all ratios
of $2J_2 /J_1$ and only the relative weights of the corresponding
susceptibilities shift with $2J_2 /J_1$.

\section{Summary and Discussion}
\label{sec:summary}

In light of the recent controversy about how best to describe the
spin dynamics of the parent FeAs compound above the magnetic ordering
temperature, we carried out a detailed magnetic susceptibility
calculation of the $J_1\!-\!J_2$ spin Hamiltonian employing classical
Monte Carlo and Schwinger boson mean field theory techniques over the
parameter range $0\le 2J_2 /J_1 \le 2$. In both treatments the
long-range magnetic ordering at finite temperature is absent and a
thorough examination of the correlated paramagnetic phase becomes
possible. In both approaches the uniform susceptibility was found to
show a largely linear temperature dependence that extends up to the
exchange energy scale $T\!\sim \! J_1$, in agreement with the
persistence of linear susceptibility up to a few hundred Kelvin in
FeAs compounds. Our calculation indicates that the linear-$T$ uniform
susceptibility behavior persists over a wider temperature range when
$2J_2 /J_1$ ratio is further removed from the critical point $2J_2
/J_1 =1$ (classical MC) or $2J_2 /J_1 =\eta_c $ (SBMFT), and is
largely independent of the degree of frustration in the spin model.

It was recently pointed out that the uniform susceptibility feature
also exists in a model based on the itinerant electron
picture\cite{korshunov}. To sharply test the validity of one model
over the other for FeAs, then, one would have to go beyond the
magnetic susceptibility issue to address other physical properties
within both approaches. It is also desirable to generalize the $J_1
\!-\! J_2$ spin Hamiltonian to the doped case where the linear
magnetic susceptibility was found to persist as well.
\\

\acknowledgments This work was supported by the Korea Science and
Engineering Foundation (KOSEF) grant funded by the Korea government
(MEST) (No. R01-2008-000-20586-0).

\appendix

\section{The susceptibility behaviors of the $J_1$ model at low
temperature}\label{J1model-sus}
The antiferromagnetic Heisenberg model with $J_2 = 0$ and $J_1 = 1$
is

\be\label{eq_app:H} H_S = \sum_{\langle ij \rangle} {\v S}_i \cdot
{\v S}_j ,\ee
has been studied in SBMFT by Auerbach and Arovas\cite{AA}, but it has
not been analyzed in detail how the structure factors and the
susceptibilities behave at low temperature for the finite system size
$N=L\times L$. In this Appendix, we will give a simple explanation
why the uniform  and staggered susceptibilities behave as
$\chi_{(0,0)}\sim A+ B T$ and $\chi_{(\pi,\pi)}\sim N/T$ at low
temperature in SBMFT for finite $N$.

The static structure factor is given by\cite{AA}
\ba\label{eq_app:S+k} S_q^+ \!=\!\sum_k\Bigl(\frac{\lambda^2
\!\!+\!\!\Delta_{2k}\Delta_{2k\!+\!q\!+\!Q}
}{E_{k}E_{k\!+\!q\!+\!Q}}\left(B_k\!+\!{1\over2}\right)
\left(B_{k\!+\!q\!+\!Q}\!+\!{1\over{2}}\right)\!-\!{1\over
4}\Bigr),\nn\ea
%
where the definition of $\Delta_{2k}$ can be found in Eq.
(\ref{eq:Delta-k}). The energy spectrum is
$E_k=\sqrt{\lambda^2-\Delta_{2k}^2}$, and the Bose-Einstein function
is $B_k=1/({e^{\beta E_k}-1})$. The minimal value of $E_k$ occurs at
$k=(0,0)$ or $k=(\pi,\pi)$, and it is convenient to introduce the gap
parameter $\delta\equiv \min(E_k)=\sqrt{\lambda^2-4\Delta_2^2}$. By
numerical calculation we find at low temperature

\be \delta =C{T^{1/2}\over L}  \label{eq:finite-size-delta}\ee
to an excellent fit and $C$ is about 3.2 $(S=1)$.

For $q=(0,0)$, the uniform structure factor becomes

\ba S_{(0,0)}^+&=& \sum_kB_k(B_{k}+1) = {1\over 4} \sum_k {1\over
\sinh^2 (\beta E_k /2) }. \nn\ea
The energy spectrum expanded near $k=(0,0)$ and $k=(\pi,\pi)$ becomes
$E_k\sim\sqrt{\delta^2 +\Delta_2^2 k^2 }$. It is useful to separate
out the two points $k=(0,0)$ and $k=(\pi,\pi)$ from the above sum and
treat the rest as an integral, which gives

\ba S_{(0,0)}^+ &\sim& {1\over 2 \sinh^2 (\beta \delta/2)} + {N\over
4\pi}\int_{1/L}^{1/a} {k dk\over{\sinh^2(\beta\Delta_2 k/2)}} \nn
& \sim &{2T^2 \over \delta^2} + { N T^2 \over \pi \Delta_2^2
}\int_{1/L}^{1/a} { dk\over k} \nn
& \sim& {2 N T\over C^2} + {N T^2 \over \pi \Delta_2^2
}\ln\left({L\over a}\right), \ea
where $1/a$ serves as an upper momentum cutoff. Based on the
relationship of the structure factor and the susceptibility in Eq.
(\ref{eq:chi-k-again}), one has

\ba \chi_{(0,0)}\sim {2 \over C^2} + { T \over \pi \Delta_2^2
}\ln\left({L\over a}\right).\ea
Indeed the low-temperature uniform susceptibility is given by the
form $A + B T$.

For $q=(\pi,\pi)$, the structure factor becomes
\ba S_{(\pi,\pi)}^+&=&\sum_k\Bigl({{\lambda^2+\Delta_{2k}^2}
\over{\lambda^2-\Delta_{2k}^2}}\left(B_{k}\!+\!{1\over2}\right)^2-
{1\over4}\Bigr).\ea
The two isolated $k$-points $k = (0,0)$ and $k=(\pi,\pi)$ contribute
to the structure factor

\ba && S_{(\pi,\pi)}^+ \sim {4\lambda^2 \over \delta^2} \times {T^2
\over \delta^2} \sim {T^2 \over \delta^4}, \ea
when we approximate $(e^{\delta /T} -1)^{-1} \approx T/\delta$.
Inserting Eq. (\ref{eq:finite-size-delta}) yields $S_{(\pi,\pi)}\sim
N^2 $ and hence the low-temperature susceptibility $\chi_{(\pi,\pi)}
\sim N/T$. The contribution from the other $k$ points is
$S_{(\pi,\pi)}^+ \sim O(N)$.


\begin{thebibliography}{99}
\bibitem{details-of-MC} The integration measure $\int d\theta \sin
\theta$ in the partition function puts an extra factor $e^{\log
(\sin \theta)}$ in the Metropolis ratio. As a result, the magnetic
moments tend to order preferentially in the $z$-direction in the MC
calculation. The magnetic susceptibilities also become
direction-dependent. One can consider, as we have done, the averaged
susceptibility $(\chi^x + \chi^y +\chi^z)/3$ after evaluating the
susceptibility for each direction individually. Instead of showing
the direction-averaged susceptibilities, we present $\chi^z$ in Fig.
\ref{fig:suscep} which shows a sharper turn and thus allows an
easier determination of $T_\mathrm{CO}$.
\end{thebibliography}
\end{document}